\documentclass[preprintnumbers]{revtex4}

\usepackage{graphicx}
\def\ie{{\it i.e.}}
\def\eg{{\it e.g.}}

\def\etal{{\it et al.}}

\def\mpl{\ifmmode \overline M_{*}\else $\overline M_{*}$\fi}
\def\to{\rightarrow}

\begin{document}
\bibliographystyle{revtex}

\preprint{CERN-OPEN-2001-077, SLAC-PUB-9030}

\title{Report of the Snowmass Subgroup on Extra Dimensions}

\author{G. P\'asztor}
\email[]{Gabriella.Pasztor@cern.ch}
\altaffiliation{On leave of absence from KFKI RMKI, Budapest, Hungary.}
\affiliation{CERN, CH-1211, Geneva 23, Switzerland}

\author{T.G. Rizzo}
\email[]{rizzo@slac.stanford.edu}
\affiliation{Stanford Linear Accelerator Center, 
Stanford University, Stanford, California 94309 USA}

\date{December 4, 2001}

\begin{abstract}
In this report we summarize the work performed at Snowmass 2001 on the physics 
of extra dimensions. We divide these analyses into the following classes: 
searches for extra dimensional phenomena, identification of specific new 
physics scenarios, studies of black hole production and non-commutative QED. 
\end{abstract}

\maketitle

\section{Searches for extra dimensions}

Many models predict the existence of additional dimensions which  
lead to new and distinct phenomenological signatures that can be searched 
for at future colliders which 
have center of mass energies in the TeV range and above. Most of the models 
in the literature fall into one of the three following broad classes: 
($i$) The large extra dimensions scenario of 
Arkani-Hamed, Dvali and Dimopoulos~(ADD)~{\cite {add}} predicts the 
emission and exchange of large Kaluza-Klein~(KK) towers of gravitons that are 
finely-spaced in mass. The emitted gravitons appear as missing energy while 
the KK tower exchange leads to contact interaction-like dimension-8 operators. 
($ii$) A second possibility includes models where the extra dimensions are of 
TeV scale in size~{\cite {anton}}. In these scenarios there are 
KK excitations of the Standard Model (SM) gauge 
(and possibly other SM) fields with masses of order TeV, which can appear as 
resonances at colliders. ($iii$) A last class of models are those with warped 
extra dimensions, such as the Randall-Sundrum Model~(RS)~{\cite {rs}}, which 
predict graviton resonances with both weak scale masses and couplings to 
matter.

\subsection{ADD Signatures in Dijets at Hadron Colliders}

The exchange of graviton towers in the ADD model can lead to substantial 
alterations in a number of $2\to 2$ processes that lead to dijet production at 
hadron colliders. Doncheski~{\cite {miked}} explored the relevant 
distributions for dijet production using a number of distinct kinematic 
variables to explore the sensitivity to ADD contributions at both the LHC and 
the Tevatron. He found that the invariant mass, the pseudorapidity ($\eta$) and 
the $\chi=exp(\eta)$ distributions were less sensitive than those associated with 
$p_t$ or $p_t^2$ in probing for graviton tower exchange. The $95\%$ CL 
exclusion reach for the Hewett cutoff scale that he obtained using the 
$p_t$ distribution was $M_s\simeq 3$ TeV at the Tevatron and 
$M_s\simeq 20$ TeV at the LHC. 

\subsection{Supersymmetric ADD}

Although the ADD model was designed to circumvent the hierarchy problem 
without the introduction of supersymmetry (SUSY), 
it may be natural that SUSY be incorporated 
into any realistic extension of this scenario. Such a possibility has been 
considered by Hewett and Sadri~{\cite {darius}}. In these extensions not only 
the graviton but also the gravitino obtain a finely grained KK tower whose 
interactions can be formulated in terms of a linearized version of 
supergravity. These authors derived the Feynman rules for this theory and 
obtained the couplings of the gravitino tower to the SUSY 
SM matter fields. As in 
the case of graviton exchange in the usual ADD model, the virtual exchange of 
gravitinos requires a cutoff when summing over the KK tower. In the ADD case 
this generates a set of effective dimension-8 operators describing the 
exchange of gravitons characterized by the scale $M_s$. 
For the gravitino case, operators of several different 
dimensions are generated, including those of dimension-6,  
so we might expect gravitino exchange effects to be large. 
In performing the sum over KK states one 
must include the shift in the propagator due to the non-zero mass of the 
zero mode that arises from SUSY breaking; however, in the leading approximation 
one finds that this zero mode mass can be neglected.

\begin{figure}[htbp]
\centerline{
\includegraphics[width=6cm,angle=90]{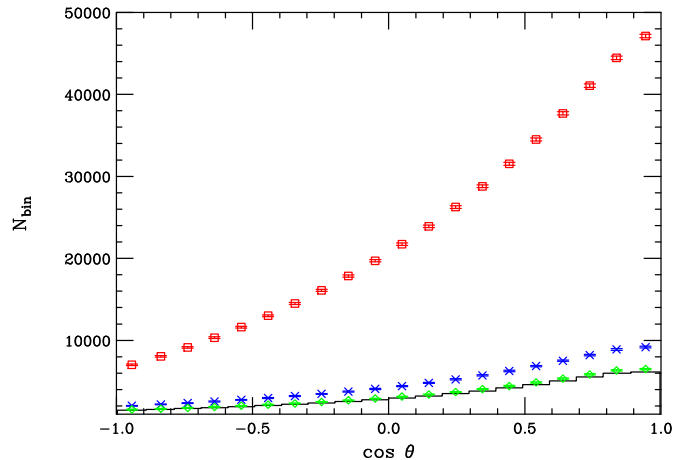}}
\vspace*{0.1cm}
\caption{Angular distribution for $e^-_Le^+ \to \tilde e_L \tilde e_R$ at a 
500 GeV linear collider with $80\%$ beam polarization for the case of six 
extra dimensions. A luminosity of 500 $fb^{-1}$ has been assumed; the 
histogram is the MSSM prediction with a bino-like LSP. The red (blue, green) 
points show the prediction with statistical errors for the SUSY ADD case 
assuming a cutoff scale of 1.5 (3, 6) TeV.} 
\label{p3-33_darius}
\end{figure}

The effects of gravitino exchange may be most noticeable in the case of 
selectron pair production at a linear collider. In addition to the MSSM 
contributions ($\gamma$ and $Z$ $s$-channel exchange and neutralino $t$-channel 
exchange) there are now contributions arising from $s$-channel gravitons and 
$t$-channel gravitinos. The use of beam polarization and the fact that the 
left and right selectrons can be separately identified helps to isolate the 
new gravitino exchange contributions to the amplitude. Figure~\ref{p3-33_darius} 
shows the size of the SUSY ADD contributions to selectron pair production in 
comparison to the usual results from the MSSM 
for one particular channel for different values of the Hewett cutoff scale. 
Here we see that values of the scale $M_s$ significantly larger than the 
collider center of mass energy can be probed via the reconstructed angular 
distribution of the selectrons. Using this selectron production channel, these 
authors showed that values of the scale $M_s$ as large as 
$\simeq 20-25 \sqrt s$ 
can be probed at linear colliders. 

\subsection{Constraints on KK Excitations in TeV-scale extra dimensions}

SM gauge bosons propagating in extra dimensions appear from the four
dimensional perspective as towers of KK states, which can be produced at 
colliders if the available energy is higher than the compactification scale 
$M_c$. Otherwise if the collision energy is insufficient for direct 
production,  they can still be observed through indirect effects. 

Cheung and  Landsberg~\cite{cheung} studied the virtual exchange of the KK 
excitations of the W, Z, $\gamma$ and $g$ bosons in high energy particle
interactions.  They performed a combined fit to the dilepton, dijet and 
top-pair production results at Tevatron, the  neutral and charged-current 
deep-inelastic scattering measurements at HERA  and the observables in dijet 
and dilepton production at LEP2, assuming that the energy scale of these 
processes are well below $M_c$ and ignoring the mixing among the KK states. 
In 
the analysis they take the leading order calculations for both the SM and 
the 
new physics processes, and to account for higher order effects they introduce a
scale factor, universal for SM and new physics, which is derived either from 
higher order SM calculations or directly from the data using regions where 
the 
contributions from KK states are expected to be vanishing. The best fit for 
$\pi^2/(3M_c^2)$ gives $-$0.29$\pm$0.09 TeV$^{-2}$ lying in the unphysical 
region, which is then turned into a lower limit of 6.8 TeV on $M_c$. 

Using the double differential cross-section $\partial^2\sigma/\partial 
M_{\ell\ell}\partial\cos\theta$ 
in the most sensitive Drell-Yan production, they estimated the reach of 
Tevatron Run II with a luminosity of  ${\cal L}$=15 fb$^{-1}$ and of LHC 
with 
${\cal L}$=100 fb$^{-1}$ to be 4.2 and 13.5 TeV, respectively.

\subsection{TeV-Scale Gauge Unification}

\begin{figure}[htbp]
\centerline{ 
\includegraphics[width=9.7cm]{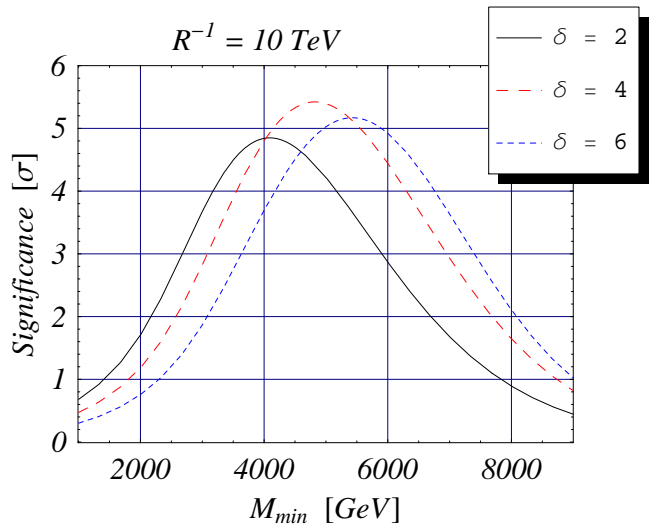}}
\caption{Deviation from the SM in units of the statistical uncertainty
in dijet production at LHC in the case of TeV-scale gauge unification
for various numbers of extra dimensions for $M_c$=10 TeV in an optimistic 
matching scenario.}
\label{fig:p3-33-unifi} 
\end{figure}

In models with TeV-scale extra dimensions the evolution of the gauge 
couplings 
are modified by power law terms. This effect can be searched for in specific 
processes which are highly sensitive to one of the gauge couplings and can be 
well measured at future colliders. 
C. Bal\'azs and B. Laforge~\cite{balazs} examined a model with TeV-scale 
extra dimensions 
with a compactification scale $M_c$ in which SM gauge bosons, 
especially the gluons, can propagate in the bulk, and gauge and Yukawa 
unifications happen slightly above the TeV range at $M_{GUT}$. They  
demonstrated that by studying dijet production at LHC, which at lowest 
order is 
sensitive to $\alpha_s^2$, one can discover the anomalous running of the 
strong 
coupling up to a common compactification scale $M_c = 5 - 10$ TeV depending 
on 
how the matching of the evolution below and above the compactification scale 
is done. To reduce the  effect of the KK excitations which compete with the 
modified running of the gauge coupling, a cut of $p_T<200$ GeV looks 
sufficient. 
The statistical significance defined as the deviation from the SM 
prediction in units of the statistical uncertainty is shown in 
Figure~\ref{fig:p3-33-unifi} for various numbers of extra dimensions 
as a function of the minimum dijet mass. 
The Tevatron Run II can observe the effect of the extra dimensions up to 
a scale of 1 TeV in the most optimistic scenario studied in the paper. 

\subsection{Extra Dimensional Signals at Far Future Colliders}

Signatures for extra dimensional models at the LHC and TeV-scale linear 
colliders are well known; many phenomenological studies and several simulation 
studies already exist. What about machines in the more distant future? Rizzo
has begun an examination of the signals for these models at both 
CLIC ($\sqrt s=3-5$ TeV and $L=1~ab^{-1}$)~{\cite {tgrclic}} and 
VLHC ($\sqrt s=175-200$ TeV and $L=0.2-1~ab^{-1}$)~{\cite {tgrvlhc}}. 

\begin{figure}[htbp]
\centerline{
\includegraphics[width=5.4cm,angle=90]{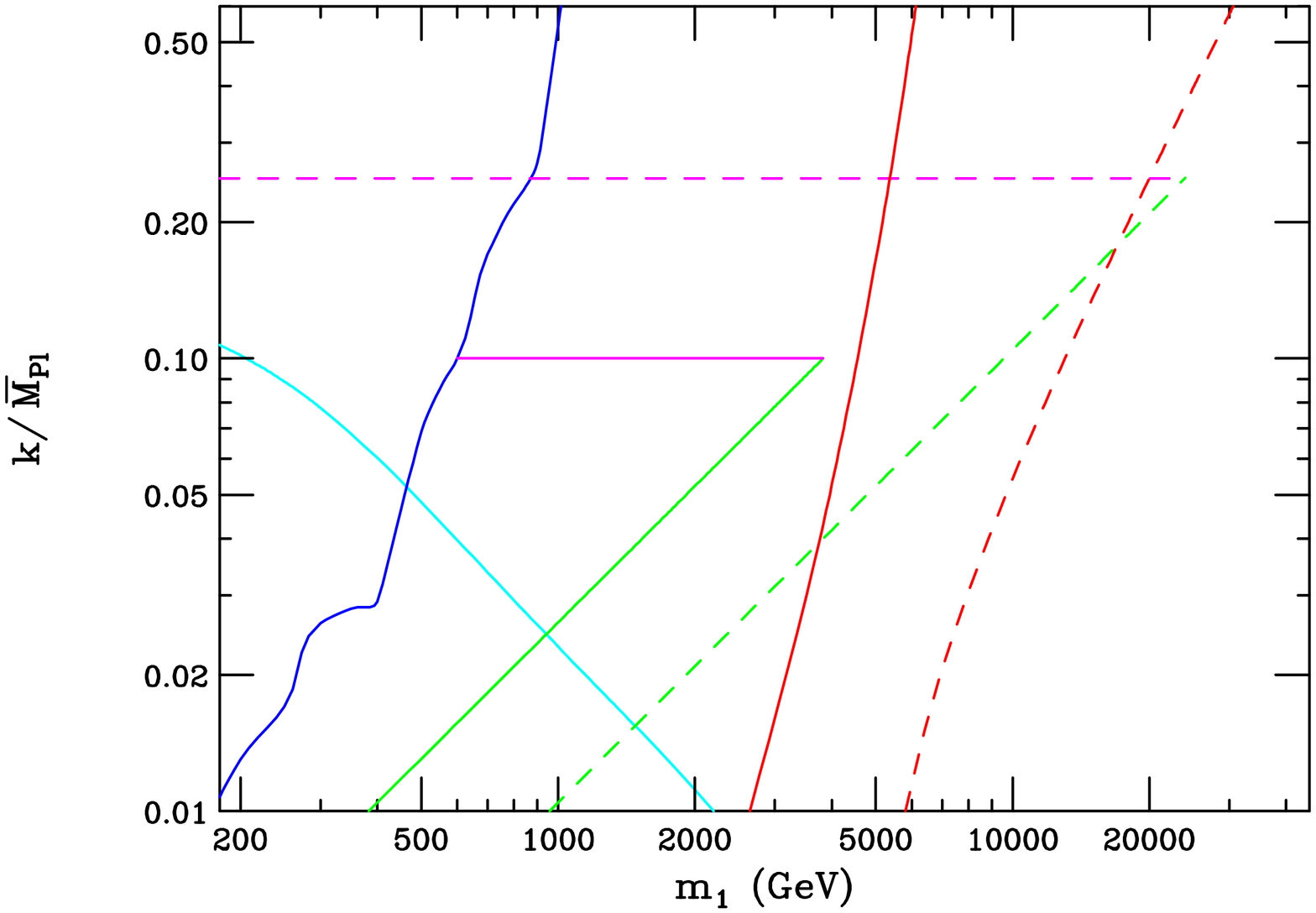}
\hspace*{5mm}
\includegraphics[width=5.4cm,angle=90]{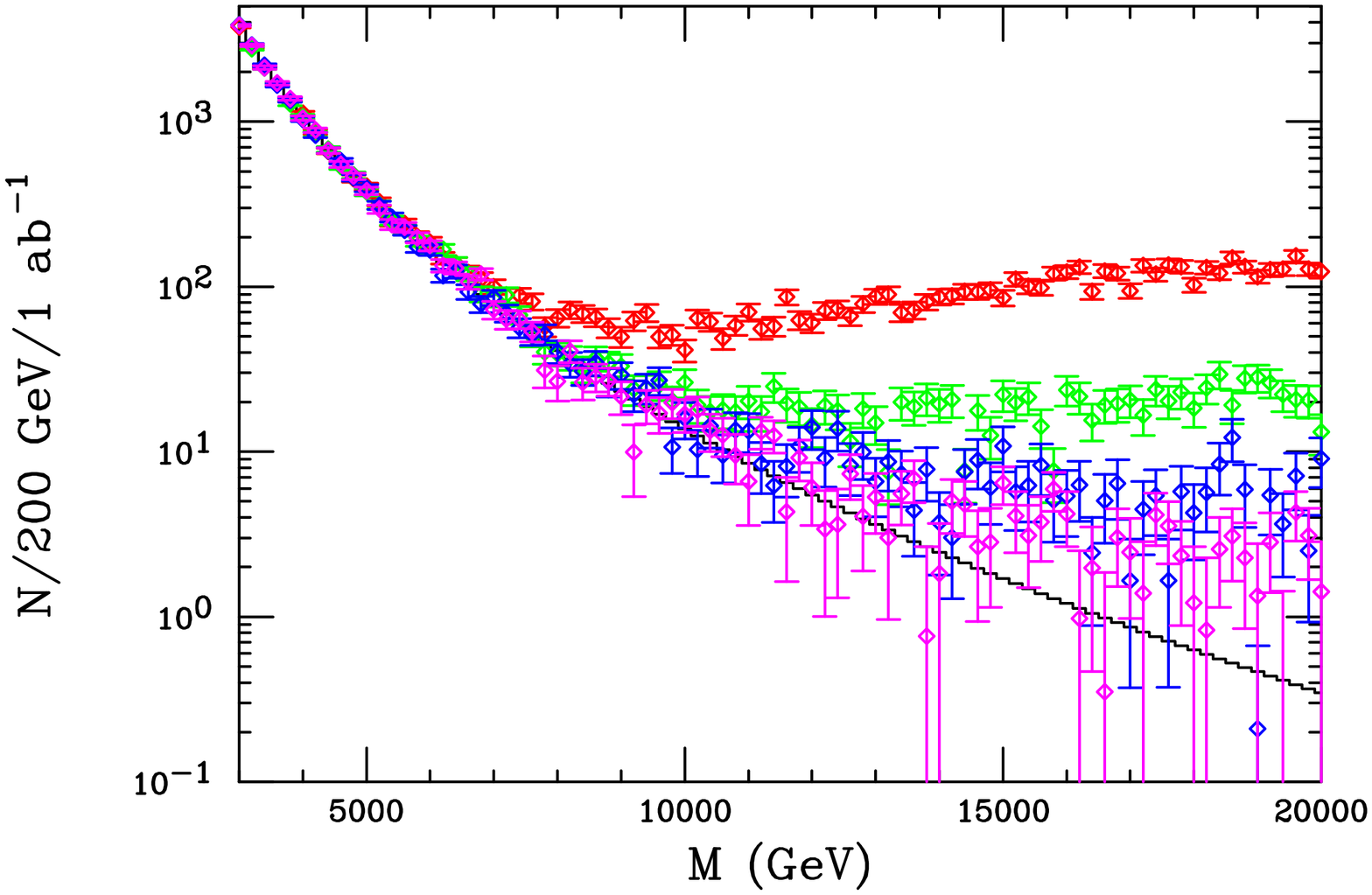}}
\vspace*{0.1cm}
\caption{(Left) Allowed regions of the RS model parameter space; $m_1$ is 
the lightest KK graviton mass. Current 
Tevatron (blue) and precision measurements (cyan) forbid regions to the left 
of their specific curves. The horizontal magenta solid (dashed) lines form the 
upper bound of the region when $c=k/\mpl=
0.10$ (0.25) while the solid (dashed) green 
curve is the corresponding lower bound when $\Lambda_\pi=10$ (25) TeV. The 
solid (dashed) red curve is the reach of the LHC (CLIC) with 100 $fb^{-1}$
($\sqrt s=5 $ TeV with 1 $ab^{-1}$). (Right) Event rate per 200 GeV mass bin 
for the Drell-Yan process as a function of the lepton pair invariant mass at a 
$\sqrt s=200$ TeV stage II VLHC. A rapidity cut $\eta_l<2.5$ on both leptons 
has been applied. The solid histogram is the SM background in both 
cases whereas the `data' points are the predictions of the ADD model. The 
red, green, blue and magenta points correspond to a Hewett cutoff scale of 
$M_s = 20, 25, 30$ or 35 TeV, respectively.}
\label{p3-33_fig1}
\end{figure}

In the case of CLIC, the highest sensitivity to both graviton exchange in 
the ADD model and gauge KK exchange in 
TeV-scale theories is to look for deviations in differential 
cross sections and asymmetries from their SM predictions due to the new 
particles. $e^+e^- \to f\bar f$ provides a set of processes of high 
sensitivity in either case from which a combined limit can be extracted. For 
a luminosity of 1 $ab^{-1}$ the reach was found to be $M_s\simeq 6\sqrt s$ for 
the Hewett cutoff scale in the ADD model and 
$M_c\simeq 16-20\sqrt s$ for the compactification scale in TeV models 
with one extra dimension compactified on 
$S_1/Z_2$. If $\sqrt s$ is sufficiently large the details of the KK 
excitation spectrum can be probed in the TeV case and the number of extra 
dimensions and the compactification manifold can be determined. Possible 
separations amongst the fermions in the extra dimensions can also be probed in 
this case. 
For the RS model there are two possibilities; if the scale of the 
graviton spectrum is low enough CLIC can be used to probe the resonance 
spectrum and the specific decay modes of the graviton including rare decays. 
If the graviton is 
somewhat more massive indirect searches can be performed in analogy to the 
ADD and TeV cases discussed above with the results as shown 
in Figure~\ref {p3-33_fig1}. In either case CLIC can greatly elucidate the 
physics of the RS model.

The VLHC was shown to have an enormous direct discovery reach for extra 
dimensional models of all types in the Drell-Yan channel. In the case of ADD 
the Drell-Yan process 
was shown to be sensitive to values of $M_s$ as large as 25$-$40 TeV through the 
production of excess events at large invariant masses as demonstrated by  
Figure~\ref {p3-33_fig1}. In the case of one extra dimension the first KK state 
of TeV-scale models could be directly observed in Drell-Yan for 
compactification scales as large as $M_c \simeq 55-60$ TeV. For models with more 
than one extra dimension, their detailed spectra could be probed in Drell-Yan 
for compactification scales in excess of 15$-$20 TeV thus allowing for a 
determination of the number of extra dimensions and their compactification 
manifolds. The production of RS graviton resonances in Drell-Yan when the SM 
fields remain on the SM brane were shown 
to be observable for masses as large as 15$-$40 TeV depending upon the value 
of the parameter $c=k/\mpl$. It was clear that if no sign of the graviton 
resonances predicted by the RS model show up by the time that CLIC/VLHC probe 
the relevant mass range then this version of the model would be excluded.

\subsection{Radions}

In the RS model, the fluctuations of the size of the extra 
dimension about its stabilized value manifest themselves as a single scalar 
field, the radion. In the RS model with a bulk scalar field 
it is expected that 
the radion is the lightest state beyond the SM fields with a mass probably 
in 
the range between ${\cal O}$(10 GeV)  and $\Lambda = e^{-kL} M_{Pl} 
\sim {\cal O}$(TeV). The couplings of the radion are in the order of 
$1/\Lambda$ and are very similar to the couplings of the SM Higgs boson, 
except 
for one important difference: due to the trace anomaly the radion directly 
couples to massless gauge bosons at one-loop. 
Moreover, in the low energy 4D effective theory the radion can mix with the 
Higgs boson, so that the physical mass eigenstates are mixtures of the radion 
and the Higgs boson. This mixing can cause important changes in the 
contribution 
to the electroweak observables compared to the no-mixing case as discussed by 
Kribs~\cite{kribs}.

\begin{figure}[htbp]
\centerline{
\includegraphics[width=5.4cm,angle=90]{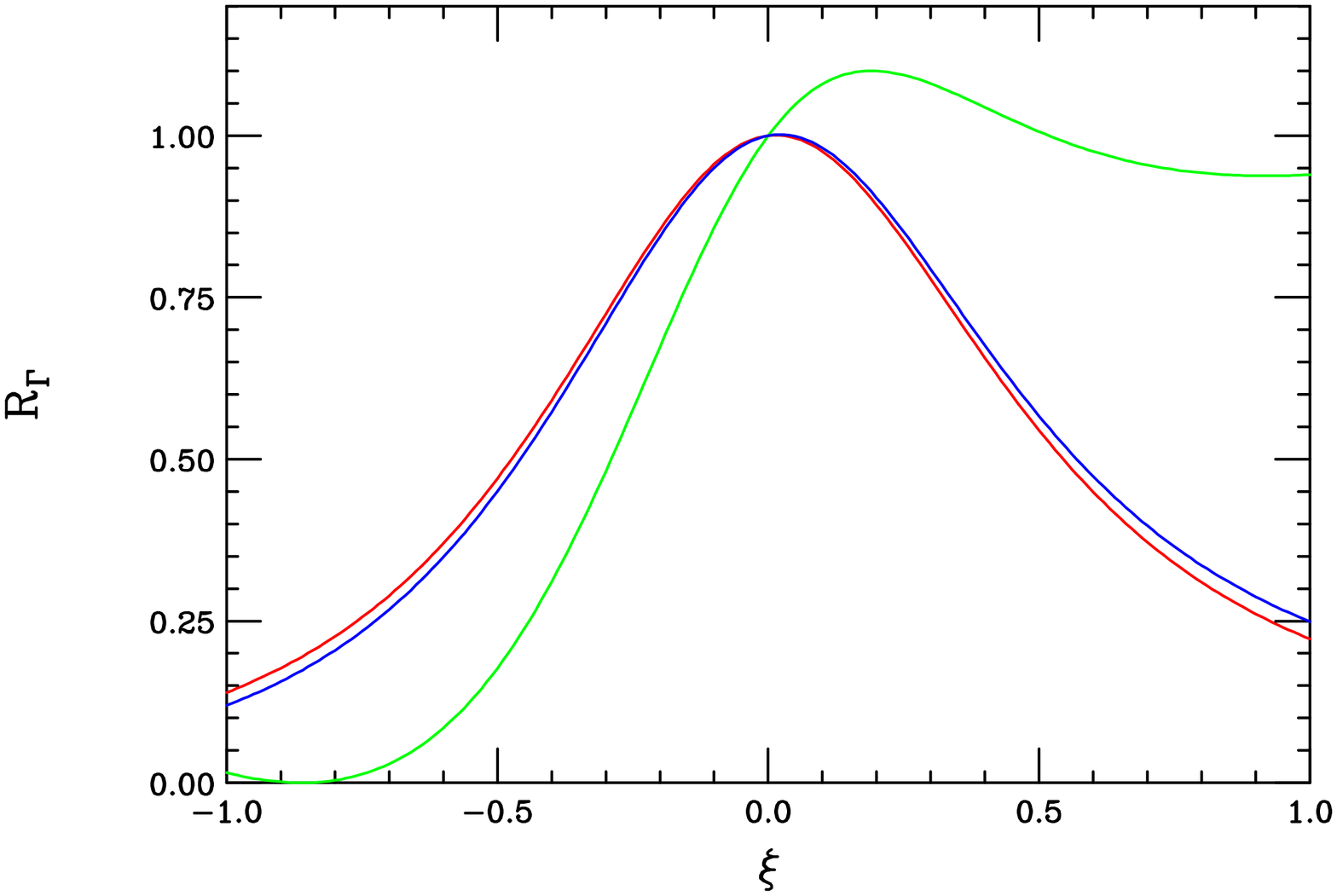}
\hspace*{5mm}
\includegraphics[width=5.4cm,angle=90]{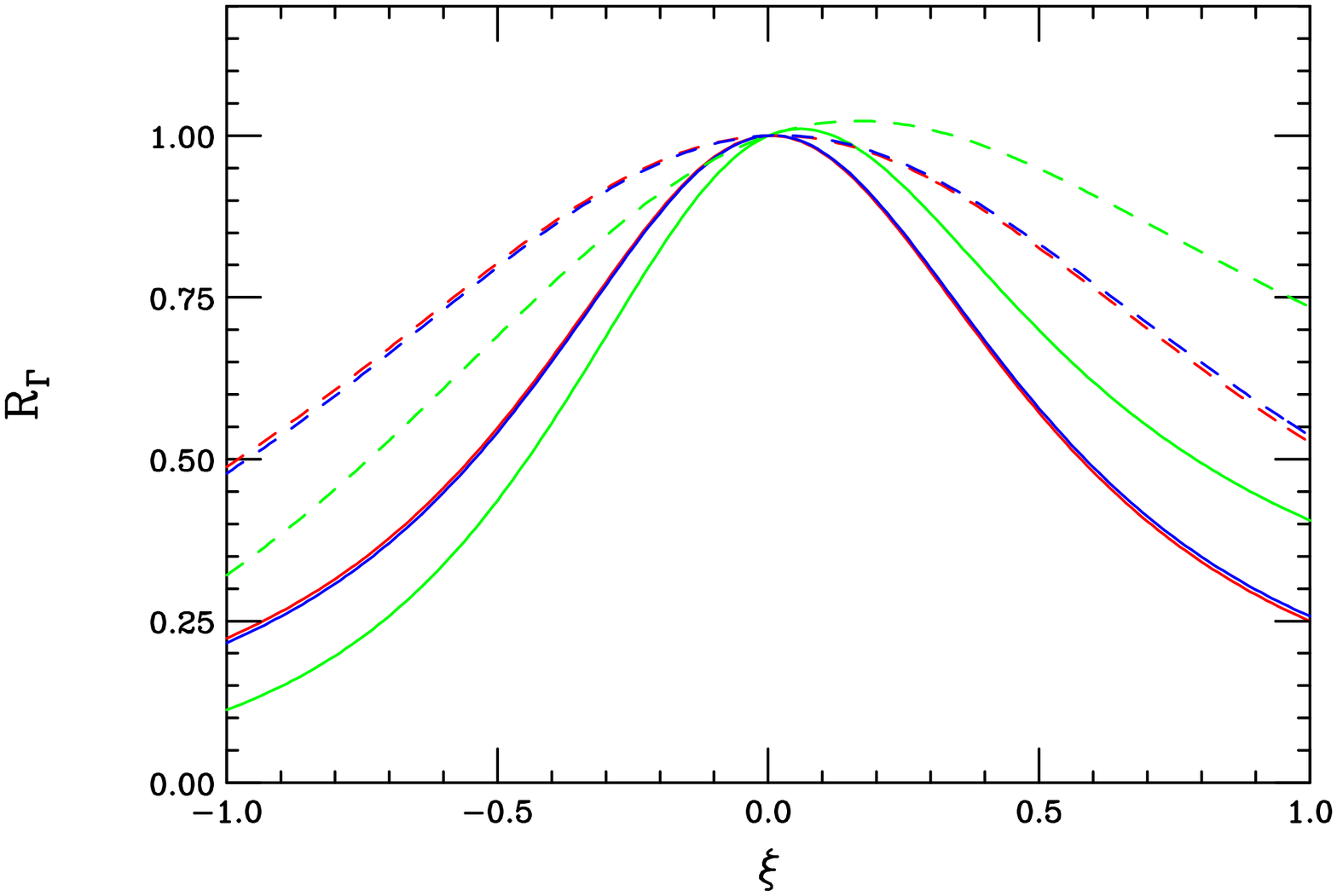}}
\vspace*{0.1cm}
\caption{Ratio of Higgs widths to their SM values, $R_\Gamma$, as a function 
of the mixing parameter $\xi$ assuming a Higgs mass of 125 GeV: red 
for fermion pairs or 
massive gauge boson pairs, green for gluons and blue for photons. In the left 
panel we assume a radion mass $m_r=300$ GeV and $v/\Lambda=0.2$. In the right 
panel the 
solid (dashed) curves are for $m_r=500$ (300) GeV and $v/\Lambda=0.2$ (0.1).}
\label{p3-33_38fig2}
\end{figure}

The mixing of the RS radion with the SM Higgs boson can lead to important 
shifts in the Higgs couplings which become apparent in the Higgs decay widths. 
This possibility was examined for these proceedings by Hewett and 
Rizzo~{\cite {jandt}}. These shifts depend upon the radion and Higgs masses, 
$m_{r,h}$, the mixing parameter $\xi$, which is expected to be of order unity, 
and the ratio of the SM Higgs vacuum expectation value,  $v$, to the RS scale $\Lambda$, which is 
of order one TeV or greater. Their results are shown in Figure~\ref{p3-33_38fig2} for 
several values of these parameters. 
Several features can be observed immediately: ($i$) the shifts in 
the widths to $\bar ff/VV$ and $\gamma \gamma$ final states are very similar,
while the corresponding shift for the $gg$ final state is quite different. 
($ii$) For relatively light radions with a low 
value of $\Lambda$ the width into the $gg$ final state can come close to 
vanishing due to a destructive interference between the contributions to the 
amplitude for values of $\xi$ near $-1$. ($iii$) Increasing the value of $m_r$ 
has less of an effect on the width shifts than does a decrease in the ratio 
${v\over {\Lambda}}$. ($iv$) Since the $VV$ and $\bar ff$ 
final states are dominant for Higgs masses in the region near 125 GeV, we 
expect that the {\it branching fractions} for these modes will be little 
influenced by radion mixing. This implies that it will be imperative to obtain 
a precise measurement of the Higgs total width in order to probe for mixing 
with radions.

\section{Identification of New Physics}

Once new physics has been discovered the next step is to identify its source. 
As is well known it is possible that many different new physics scenarios 
can lead to similar signatures at colliders. This implies that we must have a 
set of tools available to help us distinguish the various models. For example, an 
RS graviton and a $Z'$ both yield a resonance structure in the Drell-Yan 
channel at hadron colliders but they can be distinguished by the angular 
distributions of their decay products. Several analyses have addressed the 
issue of new physics identification for the case of models with extra 
dimensions for these proceedings.

\subsection{Extra Dimensions or SUSY in Missing Energy Events?}

Missing energy signatures are traditionally used to search for signs of new 
physics at colliders. The photon plus missing energy final state 
at a linear collider is particularly useful in discovering two 
classes of theories.  Theories with large extra dimensions predict the
associated production of a photon with KK excitations of the graviton 
(e$^+$e$^-$ $\rightarrow$ $\gamma$G), while in supersymmetric models 
with super-light gravitinos the photon can be produced together with a gravitino pair 
(e$^+$e$^-$ $\rightarrow$ $\gamma$ $\tilde{G} \tilde{G}$).

\begin{figure}[htbp]
\centerline{ 
\includegraphics[width=9cm]{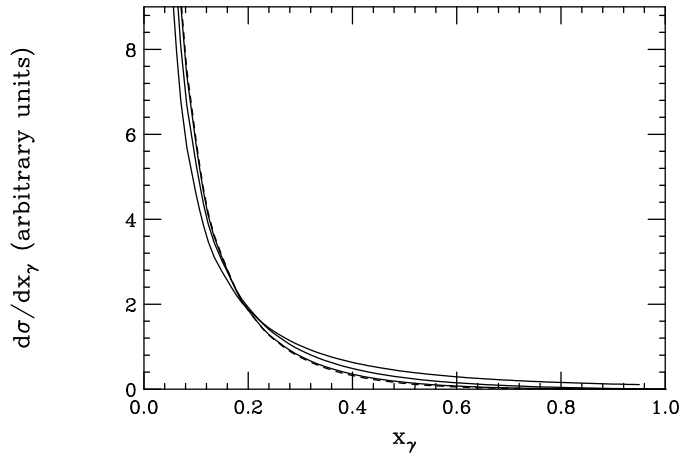}}
\vspace*{0.1cm}
\caption{Differential cross-sections of the photon plus missing energy 
final state as a function of the fractional photon 
missing energy in large extra dimensional 
theories with $n=2, 4, 6$ (solid lines) and in the super-light gravitino model 
(dashed line) with the scales of the models set so that the total 
cross-sections are equal.}
\label{fig:p3-33-emis} 
\end{figure}

Gopalakrishna, Perelstein and Wells~\cite{shri} have compared the differential 
cross-sections of the two processes as a function of $x_\gamma = 2 E_\gamma / 
\sqrt{s}$, shown in Figure~\ref{fig:p3-33-emis}, and $\cos\theta$. They found 
that a hard cut on $x_\gamma$ makes it possible to discriminate 
between supersymmetric gravitino pair-production and 
KK graviton production if the number of extra dimensions $n<6$. 
For the $n=6$ case the two models 
give very similar photon energy distributions and are indistinguishable.

Another handle to identify the models is to study the cross-section at
different center-of-mass energies. Assuming that $\sqrt{s}$ can be varied by 
5\% close to 500 GeV, the KK graviton models with lower $n$ can again be
separated  from each other and from the supersymmetric theory, but
$n=6$ gives the same scaling of the cross-section as the super-light
gravitino model. Thus one 
should look for other consequences of the models, like dimension-8 contact
interactions in extra dimensional models and the production of superpartners 
in the super-light gravitino model to identify the physics responsible for 
the production of missing energy events above the SM expectation.

\subsection{KK Resonances or $Z'$ at the LHC?}

Precision measurement constraints tell us~{\cite {anton}} that the mass of the 
lightest gauge KK excitation in models with TeV-scale extra dimensions 
is most likely greater than about 
4 TeV when SM matter fields lie at the orbifold fixed points. If this 
bound is correct then the {\it second} set of KK excitations will be too 
massive to be produced at the LHC at an observable rate. An 
essentially degenerate combination of $Z$ and $\gamma$ KK excitations has a 
sufficiently large cross section as to be observable at the LHC for masses as 
great as $\simeq 6.5-7$ TeV in the Drell-Yan channel. If such a gauge 
KK resonance is indeed observed at the LHC and 
the second excitation is too massive to be produced, 
how do we know that it is not a more conventional $Z'$? 
One argument that can be made is that in TeV KK models there will 
also be a first KK excited $W$ state which is degenerate with the first 
excitation of the $\gamma/Z$ so that a resonance should also 
appear in the charged leptons plus missing energy channel. However, there are 
many extended gauge models in the literature which predict degenerate 
$W'/Z'$ states so this observation alone will not provide proof in either 
direction. 

Rizzo~{\cite {rizzo}} has compared the Drell-Yan excitation spectrum associated 
with the production of the first KK $\gamma/Z$ excitation to that for an 
arbitrary $Z'$ with generation-independent couplings. (This is a 5 parameter 
fit assuming that the $Z'$ couples to a gauge group which commutes with weak 
isospin.) In particular, one makes 
the assumption that the KK state is actually a $Z'$ and tries to fit to the 
$Z'$'s quark and lepton couplings. A poor fit then implies that the state is 
not a $Z'$. For simplicity this analysis assumed that the 
KK excitation arose from a $S_1/Z_2$ compactification and that all the SM 
fermions were at the same fixed point ($D=0$) or that quarks and leptons were 
at opposite fixed points ($D=\pi R_c$, where $R_c$ is the compactification 
radius). Note that below the KK resonance there is a strong 
destructive (constructive) interference between the SM amplitudes and the KK 
exchanges when $D=0~(\pi R_c)$. 

The results of this analysis were as follows: 
for $D=\pi R_c$, the KK state could easily mask itself as a $Z'$ for all 
masses $M_{KK} \geq 4 $ TeV. 
For $D=0$ and relatively low values of the KK mass, 
$M_{KK} \simeq 4-5$ TeV, the $Z'$ 
hypothesis failed and the two scenarios were distinguishable. For heavier KK 
states, the statistics was so poor that model separation was lost. 
An increase in luminosity or the use of more than one channel may allow these 
results to be extended to KK states with larger masses.

\subsection{Contact Interactions at a Linear Collider}

While all present solutions to the hierarchy problem of the SM 
require the appearance of new physics at energies close to the TeV scale, the 
predicted new particles can not always be produced directly. However there may 
be a significant contribution to processes 
such as fermion or gauge boson pair production due 
to the effect of virtual states of the new heavy particles. 

If a contact interaction type correction to the Standard Model is observed,
studying its detailed properties may shed light on the fundamental physics
behind it.
P\'asztor and Perelstein~\cite{pasztor} considered several models leading to 
such indirect effects in lepton pair production at 
a 500 GeV to 1 TeV 
linear collider with the aim to determine how well one can discriminate
between them.
They studied models with large and 
TeV-scale extra dimensions and lepton compositeness~\cite{6d_contact}. 
In the large extra dimension scenario 
contributions can come from two distinct sources: the virtual exchange of KK 
excitations of the graviton or of string Regge excitations of the photon and 
the Z boson~\cite{cpp}. In the case of TeV-scale extra dimensions, 
accessible for 
the SM gauge bosons, the contact interactions arise from the exchange of the 
virtual KK excitations of the photon and the Z boson. 

It was shown that for a wide range of model parameters, measuring the lepton
pair production cross-section and angular distributions allows 
the identification of the correct candidate model. The reach in the model scale parameter for
the selection of the right model (the exclusion of {\it all} other models
considered) is close, in most cases within 5$-$15\%, to the sensitivity reach
(the exclusion of the SM) of the model. 

\begin{figure}[htbp]
\centerline{ 
\includegraphics[width=\textwidth]{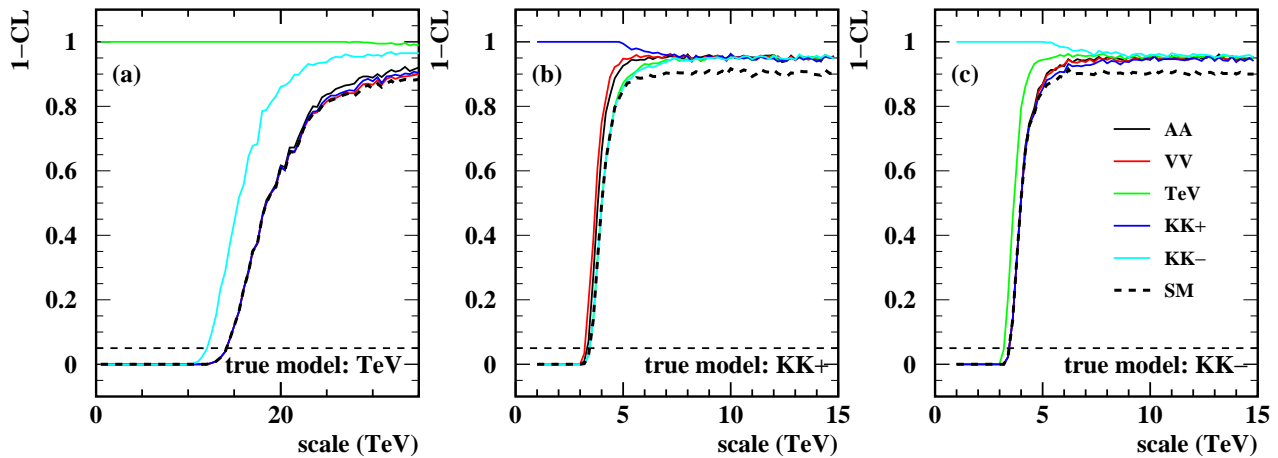}
}
\vspace*{0.1cm} 
\caption{Expected 1$-$CL at a 500
GeV linear collider with 100 fb$^{-1}$ 
luminosity after combining the three lepton pair final states, 
as a function of the scale parameter of the true model,
assumed to be (a) TeV-scale extra dimensions, large extra dimension with
dominant contribution from KK excitations of the graviton with (b) $\lambda$=+1
and (c) $\lambda$=$-$1 using the notation of Hewett, 
for various tested models. 
AA and VV stands for composite lepton models with axial-axial and vector-vector
couplings. The large extra dimension scenario with dominant contribution from 
string Regge states is omitted from this combined 
plot due to the lack of calculations
for the muon and tau pair final states.}
\label{fig:p3-33-contact}
\end{figure}

The result is demonstrated in Figure~\ref{fig:p3-33-contact}. 
Here the expected confidence level (CL) at which
the various models can be excluded is shown as a function of the scale 
parameter of the true model 
assuming that one of them (the true model) is manifested in Nature.
The combination of the electron, muon and tau pair final states proved to be 
a major tool in distinguishing the models.

\subsection{Identifying Graviton Resonances at a Linear Collider}

The production of a single spin-2 resonance at a collider may be the first 
signature for the RS model. Could it be something else? For example, some 
models predict Regge-like excitations of the SM gauge fields which also have 
spin-2~{\cite {cpp}}. However, the branching fractions of 
these particles and gravitons are 
quite different; if the various final states are accessible at a given collider,
it will allow these two possibilities to be rather easily differentiated. 
This can be done in a straightforward manner at a linear collider. 

The model by 
Dvali and co-workers~{\cite {dvali}} predicts the existence of a single 
graviton-like resonance. In this class of models the propagator of 
the graviton obtains a rather complicated structure that arises due to novel 
brane interactions, including a dimension-dependent, as 
well as energy-dependent, imaginary part and a real part which vanishes at a 
fixed value of $s$. Hence one produces an effective resonance which is in some 
sense a ``collective" mode. Since this mode is constructed 
out of a superposition of the KK states in the graviton 
tower, it has the same branching fractions as an RS graviton and thus the 
two models cannot be distinguished using such measurements. 
The Dvali \etal ~model has 
three parameters: $d \geq 2$, the number of extra dimensions, $M_d$, the 
effective resonance mass and $M_*$, the $d-$dimensional Planck scale, which is 
of the order of a few TeV. Note that in the limit $M_d \to \infty$ we 
recover the ADD model. For some values of the parameters, in particular when 
$d=5$, the effective 
resonance shape looks very much like that of a relativistic Breit-Wigner (BW) 
RS graviton. 

Rizzo~{\cite {rizzo2}} undertook a preliminary study of the line 
shapes of the resonances appearing in the RS and the Dvali \etal~models 
which are 
imagined to take place after an unfolding of the initial state radiation and 
beamstrahlung spectra; in particular one tries to fit the non-BW 
Dvali \etal ~``effective" resonance under the 
assumption that it is instead a BW RS graviton and perform a fit for the RS 
model parameter $c=k/\mpl$. A poor quality fit would thus indicate that the two 
scenarios are distinguishable. It was shown for a wide range of the other model 
parameters that the RS and Dvali \etal ~model line 
shapes could be distinguished 
provided the ratio $M_*/\sqrt s \leq 6$ and the linear collider could 
scan the region around $M_d$ with adequate luminosity.

\section{Black Hole Production at Future Colliders}

An exciting possibility in theories with extra dimensions and a low Planck 
scale is that the production rate of black holes (BH) somewhat more massive 
than $\mpl$ can be quite large at future colliders, \eg, of order 100 pb 
at the LHC~{\cite {gids}} and even larger cross sections at the VLHC; 
the actual production cross section critically depends on the BH mass, 
the exact value of $\mpl$ and the number of extra 
dimensions. Early discussions on this possibility have been extended for 
these proceedings by several authors~{\cite {sgid}} so we only review the 
essential points here. 

The basic claim of the original BH 
papers is as follows: we imagine the collision of 2 
high energy partons which are confined to a 3-brane; gravity is free to 
propagate in 
$\delta$ extra dimensions with the $4+\delta$ dimensional Planck scale 
assumed to be $\mpl \sim 1$ TeV. The curvature of the space is assumed to be 
small compared to the energy scales involved in the collision so that quantum 
gravity effects can be neglected. When these partons have a center of 
mass energy in excess of $\sim \mpl$ and the impact parameter is less than the 
Schwarzschild radius, $R_S$, associated with this center of mass energy, a 
$4+\delta$-dimensional BH 
is formed with very high efficiency. The subprocess cross section for BH 
production is thus essentially geometric, \ie, $\hat \sigma=\pi R_S^2$; 
note that $R_S$ scales as 
$[M_{BH}/\mpl^{2+\delta}]^{{1}\over {1+\delta}}$ apart from an  
overall $\delta$- and author-dependent numerical factor. (This is due to the 
different expressions for the Schwarzschild radius used by the two sets of 
authors.) This numerical factor is relatively important since it leads to a
very different $\delta$ dependence for the BH production cross section. 
This cross section expression is 
claimed to hold when $M_{BH}/\mpl$ is ``large", \ie, 
when the system can be treated semi-classically and quantum gravitational 
effects are small; one may debate just what ``large" really means, but it 
most certainly means ``at least a few". 
Voloshin has argued that an additional exponential suppression 
$\sim exp~[-C(\delta)(M_{BH}/\mpl)^{{2+\delta}\over {1+\delta}}]$, 
where $C$ is a constant,  is also present, which seriously damps the 
cross section for this process~{\cite {voloshin}} even in 
the semi-classical case. This possibility  
remains controversial at this time and strong arguments have been made on 
either side. For purposes of our discussion we will entertain 
the criticisms by  
Voloshin but remind the reader that the jury is still out on this issue. If 
they are valid one worries that the resulting cross sections for heavy BH 
will possibly be too small to be observable at the LHC; as we will see below 
this need not be the case for modest BH masses. In either case 
we anxiously await the resolution of this important issue. 

\begin{figure}[htbp]
\centerline{
\includegraphics[width=5.4cm,angle=90]{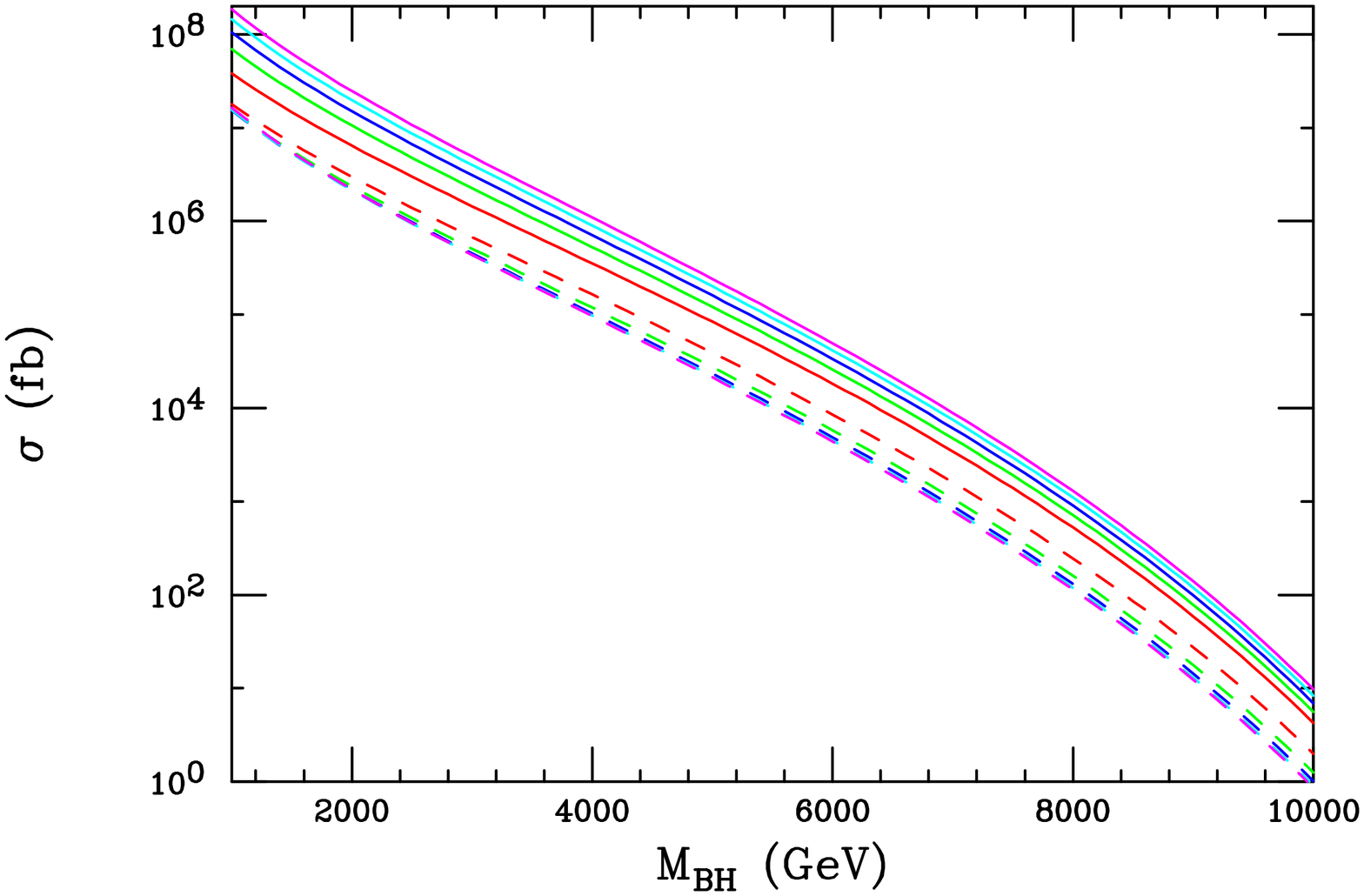}
\hspace*{5mm}
\includegraphics[width=5.4cm,angle=90]{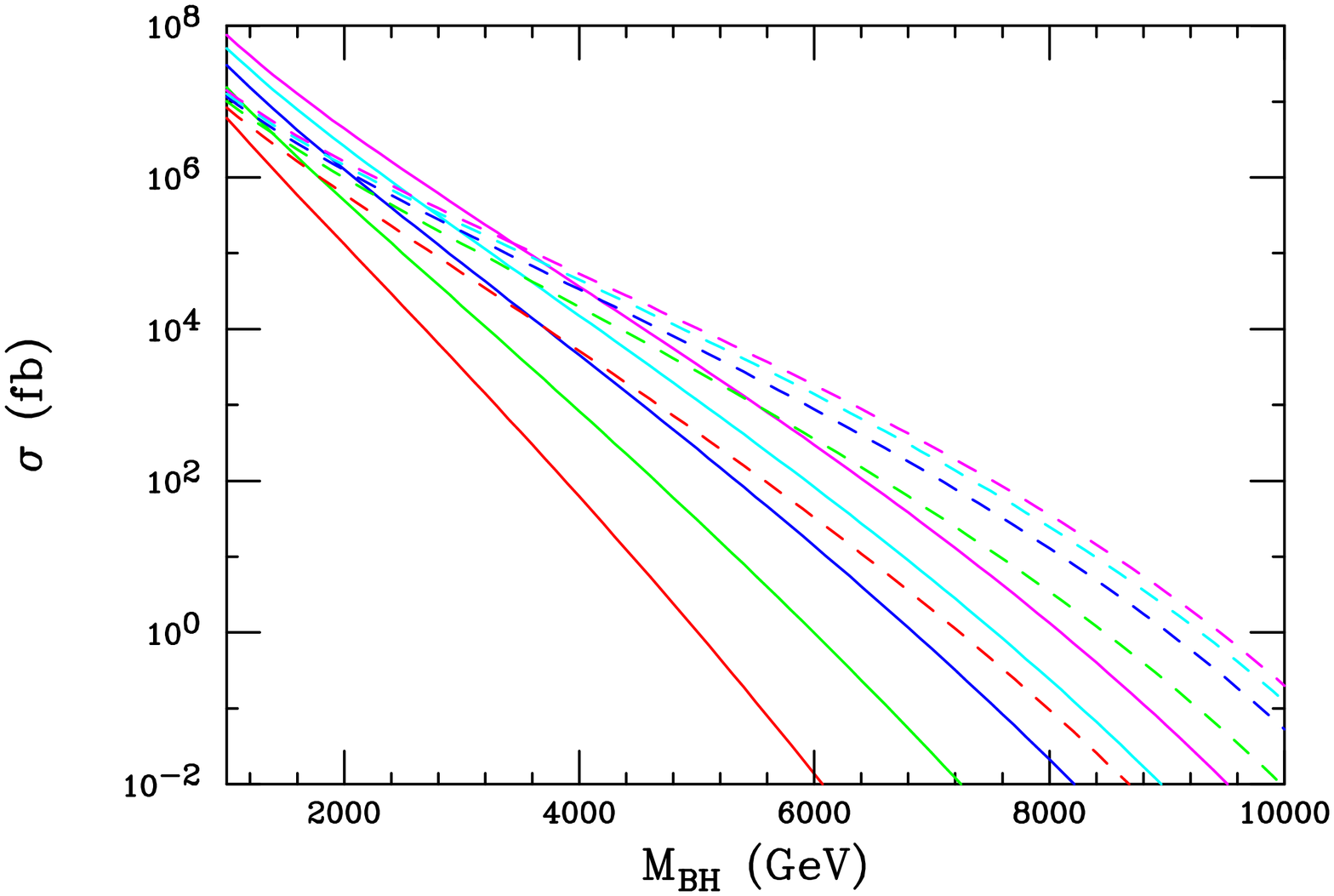}}
\vspace*{0.1cm}
\caption{(Left) Cross section for the production of BHs more massive 
than $M_{BH}$ at the LHC assuming $\mpl=1$ TeV for 
$\delta=2$ (3, 4, 5, 6) extra dimensions corresponding to the 
red (green, blue, cyan, magenta) curves. 
The solid (dashed) curves are the results 
from the work by Giddings and Thomas (Dimopoulos and Landsberg). (Right) Same 
as on the left but now including the effects of the Voloshin damping factor. 
Note that large cross sections are still possible for certain ranges of the 
parameters.}
\label{p3-33_snowholes}
\end{figure}

As can be seen in Figure~\ref{p3-33_snowholes} the rates for BH production at 
the LHC are quite large over a wide range of masses and numbers of extra 
dimensions using either set of authors' cross section expressions. Note that 
the difference between the two sets of predictions increases as $\delta$ 
increases. Once 
produced these BHs essentially decay semi-classically via Hawking radiation  
into a reasonably large number $\simeq 25$ or more final state partons in a 
highly spherical pattern. Hadrons will dominate over leptons by a factor of 
order 5$-$10 for such final states. Such signatures would not be missed at either 
hadron or lepton colliders. In addition to the above authors, Borissov and 
Lykken~{\cite {bandl}} have performed Monte Carlo studies of the decays 
of TeV scale BH which shed charge and angular momentum modeled via chemical 
potentials. These authors obtained plots of decay number densities for final 
state particles with different spins as $\delta$, $\mpl$ and $M_{BH}$ were 
varied. From this they can extract the hadronic and leptonic content of BH 
decays as well as the contributions due to photons, gravitons and Higgs 
bosons. We note that an alternative decay scenario has been advocated by 
Casadio and Harms~{\cite {casa}}. 

Note that Figure~\ref{p3-33_snowholes} also shows the effects of 
the suppression predicted by Voloshin using the authors' respective 
expressions for $R_S$~{\cite {rizzo3}}; 
here we make the important observation that at least for some 
range of parameters BHs will still be produced at significant rates to be 
observable at the LHC {\it even if the Voloshin suppression is active}.

\section{Non-commutative QED}

If the Planck scale is indeed of order of a few TeV, other possible stringy 
effects may be observable at future colliders. One possibility is that 
space-time is non-commutative (NC), \ie, 
$[x_\mu,x_\nu]=i\theta_{\mu\nu}=ic_{\mu\nu}/\Lambda_{NC}^2$, where 
the $c_{\mu\nu}$ is a set of parameters of order unity and 
$\Lambda_{NC} \sim $ a few TeV. NC effects thus only appear as the TeV scale 
is approached, generally appearing as dimension-8 operators. 
NC theories are Lorentz 
violating (but CPT conserving) since we can decompose the frame-independent 
quantity $c_{\mu\nu}$ into two 3-vectors (in analogy with the $E$ and $B$ 
fields) which point in some fixed but 
{\it a priori} unknown preferred directions. 
NC QED provides a potential experimental testing ground for such theories and 
has interesting collider signatures~{\cite {ncpheno}}. 
NC QED differs from ordinary QED in 
several ways: ($i$) the $ee\gamma$ vertex picks up a Lorentz violating 
phase factor which is dependent 
upon the electron momenta and the components of $\theta_{\mu \nu}$; ($ii$) the 
NC theory predicts trilinear and quartic couplings between the photons that 
are, to leading order, linear and quadratic in the parameters 
$\theta_{\mu\nu}$ and are kinematics dependent; ($iii$) only the charges 
$Q=0,\pm 1$ are allowed by gauge invariance in NC QED and thus quarks cannot 
be treated in the theory as it presently exists and an extension to the full 
NC SM is required. The hallmark signal at 
colliders for NC QED is the appearance of an azimuthally-dependent cross 
section in $2\to 2$ processes; the azimuthal dependence arises from the 
existence of preferred directions. 

\begin{figure}[htbp]
\vspace*{0.1cm}
\centerline{
\includegraphics[width=6cm,angle=-90]{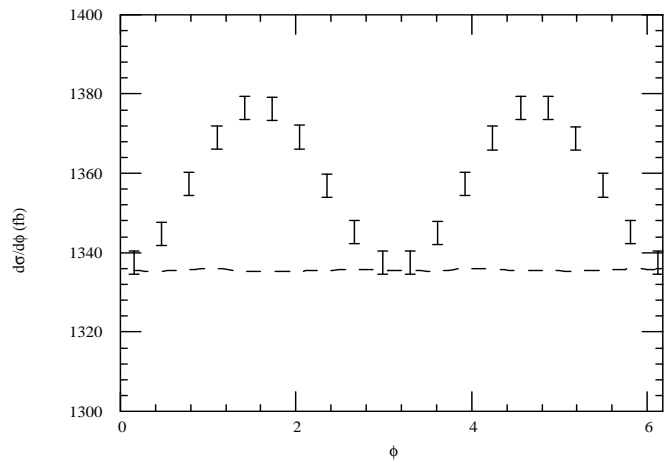}}
\vspace*{0.1cm}
\caption{Binned azimuthal cross section dependence for the
Compton scattering of 
unpolarized photons at a 500 GeV linear collider assuming $\Lambda_{NC}=500$ 
GeV and an integrated luminosity of 500 $fb^{-1}$. The dashed line in the 
SM expectation.}
\label{p3-33_steve}
\end{figure}

Godfrey and Doncheski~{\cite {gdnc}} have examined the two processes 
$\gamma \gamma \to e^+e^-$ and $\gamma e \to \gamma e$ for the effects of NC 
at linear colliders in the 0.5-8 TeV energy range. Their analysis assumed an 
integrated luminosity of 500 $fb^{-1}$ and sets a series of lower bounds on 
the scale $\Lambda_{NC}$. (Note that while $\gamma \gamma$ probes 
space-time (\ie, $c_{0i}$) NC, $\gamma e$ was shown to probe both 
space-space (\ie, $c_{ij}$) and space-time NC.) 
Figure~\ref{p3-33_steve} from~{\cite {gdnc}} shows the azimuthal dependence 
of the Compton scattering 
cross section at a 500 GeV collider when $\Lambda_{NC}$=500 GeV 
for a specific orientation of the $c_{0i}$ and $c_{ij}$ vectors relative to 
the beam axis; we observe that the deviations are quite statistically 
significant. For the $\gamma \gamma$ process 
with $c_{0i}$ perpendicular to the beam axis the authors' 
$95\%$ CL lower bounds on 
$\Lambda_{NC}$ are in the range $\simeq 0.42-0.52 \sqrt s$ while bounds as 
high as $\simeq 1.5\sqrt s$ are obtained for the $\gamma e$ process. 
In the $\gamma \gamma$ case the size of the NC effect was shown to 
depend  only upon the cosine 
of the angle between the beam axis and the NC direction while in the Compton 
case the angular dependence was found to be somewhat more complex depending 
on two orientation angles. 

\section*{Acknowledgement} 

G.P. was partially supported in Snowmass by a DPF Snowmass Fellowship and by
the Hungarian Scientific Research Fund under the contract numbers OTKA T029264
and T029328.

\def\MPL #1 #2 #3 {Mod. Phys. Lett. {\bf#1},\ #2 (#3)}
\def\NPB #1 #2 #3 {Nucl. Phys. {\bf#1},\ #2 (#3)}
\def\PLB #1 #2 #3 {Phys. Lett. {\bf#1},\ #2 (#3)}
\def\PR #1 #2 #3 {Phys. Rep. {\bf#1},\ #2 (#3)}
\def\PRD #1 #2 #3 {Phys. Rev. {\bf#1},\ #2 (#3)}
\def\PRL #1 #2 #3 {Phys. Rev. Lett. {\bf#1},\ #2 (#3)}
\def\RMP #1 #2 #3 {Rev. Mod. Phys. {\bf#1},\ #2 (#3)}
\def\NIM #1 #2 #3 {Nucl. Inst. Meth. {\bf#1},\ #2 (#3)}
\def\ZPC #1 #2 #3 {Z. Phys. {\bf#1},\ #2 (#3)}
\def\EJPC #1 #2 #3 {E. Phys. J. {\bf#1},\ #2 (#3)}
\def\IJMP #1 #2 #3 {Int. J. Mod. Phys. {\bf#1},\ #2 (#3)}
\def\JHEP #1 #2 #3 {J. High En. Phys. {\bf#1},\ #2 (#3)}

\end{document}